\documentclass[sigconf,nonacm]{acmart}

\usepackage{listings} 
\usepackage{xcolor}
\usepackage{mdframed} 
\usepackage{tikz}
\usepackage{longtable}
\usetikzlibrary{arrows.meta, positioning}
\usepackage{todonotes}
\definecolor{codered}{rgb}{0.6,0,0}
\definecolor{codegreen}{rgb}{0,0.6,0}
\definecolor{codegray}{rgb}{0.5,0.5,0.5}
\definecolor{codepurple}{rgb}{0.58,0,0.82}
\definecolor{backcolour}{rgb}{0.95,0.95,0.92}
\lstdefinestyle{mystyle}{
    backgroundcolor=\color{backcolour},   
    commentstyle=\color{codegreen},
    keywordstyle=\color{magenta},
    numberstyle=\tiny\color{codegray},
    stringstyle=\color{codepurple},
    basicstyle=\ttfamily\footnotesize,
    breakatwhitespace=false,         
    breaklines=true,                 
    captionpos=b,                    
    keepspaces=true,                 
    numbersep=5pt,                  
    showspaces=false,                
    showstringspaces=false,
    showtabs=false,                  
    tabsize=2
}
\usepackage{enumitem}
\usepackage{xcolor}
\usepackage{tikz}
\definecolor{customMagenta}{HTML}{E52CCF}

\usepackage{tabularx}   
\usepackage{booktabs}  
\usepackage[table]{xcolor} 
\usepackage{array} 
\AtBeginDocument{%
  \providecommand\BibTeX{{%
    \normalfont B\kern-0.45em{\scshape i\kern-0.25em b}\kern-0.8em\TeX}}}
\makeatletter
\newcommand{\acceptedversionnotice}{%
  \copyright~Jessica Hutchison, Ian Tyler Applebaum, Kenneth Angelikas, Kush Rakesh Patel, Phuoc Nguyen, Antonio Lazaro, Nicholas Rucinski, Rahad Arman Nabid, and Stephen MacNeil 2026. This is the author's version of the work. It is posted here for your personal use. Not for redistribution. The definitive Version of Record was published in Proceedings of the 31st ACM Conference on Innovation and Technology in Computer Science Education (ITiCSE 2026), \url{http://dx.doi.org/10.1145/3803400.3809394}.%
}
\def\@authorfont{\Large}
\def\@affiliationfont{\normalsize\normalfont}
\setcopyright{cc}
\gdef\@copyrightpermission{{\fontsize{6}{7}\selectfont\acceptedversionnotice}}
\makeatother

\begin{document}


\title[To Tab or Not to Tab]{To Tab or Not to Tab: Measuring Critical Engagement in AI Code Completion Tools Using Behavioral Signals and Attention Checks}

\author{Jessica Hutchison}
\affiliation{%
  \institution{Temple University}
  \city{Philadelphia}
  \country{USA}
}
\email{jessica.hutchison@temple.edu}
\orcid{0009-0007-5775-9631}

\author{Ian Tyler Applebaum}
\affiliation{%
  \institution{Temple University}
  \city{Philadelphia}
  \country{USA}
}
\email{ian.tyler@temple.edu}
\orcid{0009-0002-0704-9379}

\author{Kenneth	Angelikas}
\affiliation{%
\institution{Temple University}
  \city{Philadelphia}
  \country{USA}
}
\email{kenneth.angelikas@temple.edu}
\orcid{0009-0001-1597-3513}

\author{Kush Rakesh Patel}
\affiliation{%
  \institution{Temple University}
  \city{Philadelphia}
  \country{USA}
}
\email{tur46917@temple.edu}
\orcid{0009-0008-0356-3294}

\author{Phuoc Nguyen}
\affiliation{%
  \institution{Temple University}
  \city{Philadelphia}
  \country{USA}
}
\email{jaime.nguyen@temple.edu}
\orcid{0009-0009-5780-7060}

\author{Antonio	Lazaro}
\affiliation{%
  \institution{Temple University}
  \city{Philadelphia}
  \country{USA}
}
\email{tut08369@temple.edu}
\orcid{0009-0005-6520-6899}

\author{Nicholas Rucinski} 
\affiliation{%
  \institution{Temple University}
  \city{Philadelphia}
  \country{USA}
}
\email{nicholas.rucinski@temple.edu}
\orcid{0009-0005-6780-8017}

\author{Rahad Arman Nabid}
\affiliation{%
  \institution{Temple University}
  \city{Philadelphia}
  \country{USA}
}
\email{rahad.arman.nabid@temple.edu}
\orcid{0000-0002-6562-6595}

\author{Stephen MacNeil}
\affiliation{%
  \institution{Temple University}
  \city{Philadelphia}
  \country{USA}
}
\email{stephen.macneil@temple.edu}
\orcid{0000-0003-2781-6619}

\renewcommand{\shortauthors}{Jessica Hutchison et al.}

\begin{abstract}

AI code completion tools, such as Github Copilot, provide students with code suggestions to help them write programs. However, recent qualitative studies suggest that students fail to critically evaluate these suggestions. We present Clover, a code completion tool that logs students' interactions with code suggestions and additionally offers \textit{attention checks} to probe reflective engagement during programming tasks. We also develop a taxonomy of behavioral interaction metrics for AI-assisted programming, informed by literature. We analyzed relationships between interaction patterns, engagement with attention checks, and task performance. We observed that higher rates of \textit{tab accept} were associated with lower attention check performance, while increased \textit{dwell time} was associated with higher attention check performance. We conclude by discussing how programming process data and attention checks might support  reflective engagement in AI-assisted programming.

\end{abstract}


\maketitle

\section{Introduction} 

AI-assisted programming is rapidly changing how software is written, with developers increasingly working alongside systems that generate code suggestions in real time. AI code completion tools, such as GitHub Copilot and Cursor, are increasingly integrated into professional workflows and, more recently, into computing courses as well~\cite{vadaparty2024cs1, porter2024learn}. Although student adoption of code assistants has lagged behind general purpose AI tools such as ChatGPT, recent surveys show their use by computing students is steadily increasing~\cite{hou2025evolving, keuning2024perceptions, ko2025rethinking, prather2023robots, smith2024early}. As these tools become standard, fundamental questions arise about how students interact with AI code suggestions and how to meaningfully measure these interactions. 

Despite growing adoption, understanding students' interactions with AI code completion tools remains a challenge. Existing studies rely on qualitative methods including think-aloud protocols~\cite{barke2023grounded, wermelinger2023GitHub}, screen recordings~\cite{prather2024widening, shihab2025effects}, screenshots ~\cite{menon2023exploring}, and process journals ~\cite{shah2025students}. While these approaches provide rich insights they are difficult to scale and offer limited visibility into fine-grained real-time interactions with code suggestions. 

This measurement gap is important given emerging evidence that AI code assistants can impact learning behaviors and outcomes. Prior work has suggested that code completion tools can mislead students and disrupt their metacognitive processes~\cite{prather2024widening, park2025evaluating}. They can also lead to unproductive debugging `rabbit holes'~\cite{vaithilingam2022expectation}. These negative impacts tend to disproportionately affect students with less experience and lower self efficacy~\cite{bernstein2025beyond, prather2024beyond, hou2024effects, zastudil2023generative, margulieux2024self}. Therefore, it is critical to develop metrics and methods for evaluating how students interact with these tools.   

To address this gap, we introduce Clover, a code completion tool that instruments student interactions with AI-generated suggestions and introduces lightweight \textbf{attention checks} to probe reflective engagement during programming tasks. To inform the metrics Clover tracks, we developed a comprehensive Taxonomy of Interaction Behaviors with AI Code Suggestions guided by metrics used across prior literature~\cite{prather2023weird, prather2024widening, barke2023grounded, vaithilingam2022expectation, kazemitabaar2024selfpaced}. We then conducted a lab study with Clover in a CS1 course and analyze relationships between interaction patterns, attention check performance, and task performance to investigate the following research questions:

\begin{itemize}

    \item [\textbf{RQ1:}] How does CS1 students' usage of a code completion tool relate to their task performance?
    \item [\textbf{RQ2:}] How does CS1 students' engagement with attention checks relate to their task performance?
\end{itemize}

\noindent In this paper, we make the following contributions: 
\begin{itemize}
    \item \textbf{Clover, an instrumented code completion tool} that logs fine-grained interactions with AI-generated suggestions and integrates \textit{attention checks} to probe reflective engagement.
    \item \textbf{A taxonomy of interactions with AI code suggestions} for AI-assisted programming synthesized from prior work. 
    \item \textbf{An empirical study} demonstrating how interaction patterns relate to engagement and task performance.

\end{itemize}

\section{Related Work}

Programming Process Data (PPD) captures how students interact with the integrated development environment while coding. PPD can be collected at varying levels of granularity, from coarse-grained artifacts, such as final submissions, to fine-grained logs of individual keystrokes~\cite{edwards2025opportunities, leinonen2019keystroke}. PPD has been used to study how novice programmers code, identify sources of struggle, and support early intervention or integrity monitoring~\cite{price2020progsnap2, edwards2025opportunities, leinonen2019keystroke, gao2021early, brown2014investigating, karol2025koala}. However, this prior work has been focused on traditional programming environments without AI assistance.

The introduction of AI code completion tools have surfaced diverse  interaction patterns and complex impacts on learning. Prior work suggests that programmers interact with AI tools differently depending on their goals and experience. For example, \citet{barke2023grounded} conducted one of the earliest studies of how professional programmers interact with Github Copilot identifying two modes: \textit{exploration} where programmers use Copilot to explore options and \textit{acceleration} where the programmer knows what to do and uses Copilot to get there faster. While this shows the flexibility of code assistants for experts, subsequent work suggests that novices may struggle to use these tools effectively. 

Based on an eye tracking study of students using Github Copilot, ~\citet{prather2024widening} found that less prepared students were often misled by AI suggestions and struggled to differentiate good suggestions from bad ones. This reflected a broader trend of unequal learning benefits associated with AI use~\cite{hou2024effects, prather2023robots, bernstein2025beyond, prather2024widening, zastudil2023generative}. Another study by \citet{vaithilingam2022expectation} showed how students could get stuck in `debugging rabbit holes' when using code generated by AI code assistants. Additional issues include students experiencing an `illusion of competence'~\cite{prather2024widening} where students have difficulty identifying gaps in understanding~\cite{prather2024widening, margulieux2024self}. 

However, understanding these interaction patterns remains challenging. Most studies have used qualitative methods such as screen capture~\cite{shihab2025effects}, eye-tracking ~\cite{prather2024widening}, or manual observations~\cite{shihab2025effects}. While these approaches provide valuable insights, they are labor-intensive and difficult to scale. PPD offers a complementary, scalable approach for analyzing behaviors such as accepting or rejecting suggestions across larger populations. However, existing PPD approaches have not been fully adapted to capture these interaction patterns.

\section{Methods} 

\subsection{Clover Design and Implementation}
Clover is an integrated development environment (IDE) extension that uses the Gemini 3 Large Language Model to provide real-time, single-line code suggestions at the cursor position based on code already in the editor. While programming, students can accept these code suggestions or write their own code manually. To capture how students engage with suggestions, Clover logs students' interactions, which are described in detail in Section~\ref{sec:data-collection}.

\subsubsection{Instrumenting the Copilot Experience}
To maximize ecological validity, Clover was implemented as a Visual Studio Code extension, mirroring the interface and functionality of widely-adopted tools like GitHub Copilot. We kept familiar interaction models, such as \textit{Tab-to-accept}, with code suggestions appearing automatically when a user pauses during typing. By maintaining this user experience, students likely behave as they would in a real-world programming environment, allowing us to collect authentic behavioral data. 

\subsubsection{Single-Line Code Suggestions}
Large blocks of AI code suggestions can be difficult to understand and debug~\cite{vaithilingam2022expectation}, so Clover provides single-line code suggestions to reduce cognitive load and provide granular detection of behavioral metrics.

\subsubsection{Attention Checks} 

Over-reliance on AI code completion tools can occur when users default to \textit{fast thinking}, accepting suggestions without the critical evaluation required to determine their correctness~\cite{evans2008dual, kahneman2011thinking, harbarth2025over}. In order to detect students' critical engagement with AI code suggestions, we intentionally inject a controlled number of suggestions that are semantically inconsistent with the student's immediate coding goal. We say `inconsistent' instead of `incorrect' because these \textit{attention checks} could still produce functional code through later corrections~\cite{macneil2025fostering}. For example, a student creating a counter might be shown \verb|count--|, when what they need is \verb|count++|. If the student accepts this suggestion, they \textit{fail} the \textit{attention check}; and if the student rejects the suggestion, they \textit{pass}. These \textit{attention checks} are intended to be similar to misleading suggestions and hallucinations in tools like Github Copilot~\cite{macneil2024synthetic}, but are deliberate instead of incidental. This makes them useful in determining whether the student is paying attention. This is similar to Harbarth's method of giving incorrect routes as AI suggestions in a driving context~\cite{harbarth2025over}.

\subsection{In-Person Classroom Study}

We conducted an initial study of Clover which was approved by the Institutional Review Board (IRB) at Temple University. 

\subsubsection{Course Context}

This study was conducted in a first-year programming course that serves as the second course in the introductory programming sequence. The in-person course is taught in Java and focuses on object-oriented programming and data structures. The course contains mix of students in computer science and related majors. The course features a weekly programming lab that runs for 110 minutes. Students attend lab in four separate sections. 

The course instructor typically prohibits the use of generative AI, and instead encourages students to get help from peers or instructional staff; however, they made an exception for this lab with the intention to give their students exposure to AI tools.

\subsubsection{Participants} 
We recruited students from the course, offering extra credit as compensation. Students who opted not to participate could receive equivalent credit by completing the assignment without Clover. 56 students spread across four different sections of this course consented and participated in every part of the study. Details about the study were presented verbally alongside a digital consent form. Students were made aware that their participation was entirely voluntary. Data was anonymized using unique participant IDs and stored securely in a password-protected database. One participant was removed because their average dwell time was three standard deviations from the mean, resulting in 55 students included in the analysis.

\subsubsection{Study Procedures}

The study took place simultaneously across the four lab sections with each following the same procedure: 1) 15 minutes were allocated to explain the study and for participants to complete the consent form, 2) students were introduced to the tool and given a warm-up problem to become familiar with the system for another 15 minutes, 3) students accessed our system through URLs to GitHub Codespaces, a browser based VSCode, with Clover preinstalled, 4)  participants were assigned the rainfall problem to complete within a 60-minute time frame. During the task, Clover intermittently produced deliberately incorrect suggestions as attention checks. Participants were not informed of this behavior in advance because this awareness would have undermined the purpose of the manipulation. This use of deception was reviewed and approved by the Institutional Review Board.

\subsection{Data Collection}
\label{sec:data-collection}

We instrumented Clover to log fine-grained interaction events as participants worked on a programming task. These events capture changes in the code state that occur when a suggestion is generated, displayed, or acted upon, allowing us to observe both what decisions participants make and how those decisions are reached. Events are stored with a timestamp of when the event occurred. 

\subsection{Taxonomy of Interaction Metrics}
\label{sec:taxonomy}

Prior work has characterized a range of interaction patterns with AI code completion tools, but these characterizations are fragmented across studies.
For example, Prather et al. conducted a study on Github Copilot use by CS1 students and identified slow accept, backtrack, and adapt behaviors~\cite{prather2023weird}. These behaviors could more broadly be attributed to two novice programmer types, drifters and shepherds. Similarly, Barke et al. observed two modes of interaction, acceleration and exploration~\cite{barke2023grounded}. Other studies show that students respond to incorrect AI suggestions by either attempting to repair accepted code or deleting it entirely ~\cite{vaithilingam2022expectation}. Shihab et al. displays how high efficiency in completing the `brownfield' tasks can mask levels of comprehension. Accordingly, we include efficiency and effectiveness metrics to understand participants' levels of comprehension~\cite{shihab2025effects}.

These studies present overlapping but inconsistently defined behavioral constructs, making it difficult to compare results across systems and contexts. In some cases, similar behaviors are labeled differently, while in others, identical labels refer to different constructs. To address this, we synthesize prior work into a unified taxonomy of behavioral metrics for AI code completion tools.

\subsubsection{Suggestion Generation Events} Clover logs the following three sequential events leading up to displaying a suggestion:

\begin{itemize}
    \item[\textbf{GAC}] \textit{Generate\_Attention\_Check}: Generating a deliberately misleading suggestion.
    
    \item[\textbf{GS}] \textit{Generate\_Suggestion}: Generating a code suggestion that is \textit{not an attention check}. 
    
    \item[\textbf{SS}] \textit{Suggestion\_Shown}: Visually presenting a new code suggestion or attention check \textbf{(AC)} after it is generated. 
\end{itemize}

\subsubsection{Acceptance Events} We consider the following two instances to indicate a suggestion was accepted: 

\begin{itemize}

    \item[\textbf{TA}] \textit{Tab\_Accept}: Pressing `tab' to accept a suggestion, without immediate modification or deletion. 
    \item[\textbf{SA}] \textit{Slow\_Accept}: Typing out a suggestion character by character without clicking tab. This behavior was first observed qualitatively in Prather's study of GitHub Copilot~\cite{prather2023weird}. 
\end{itemize}

\subsubsection{Rejection and Revision Events} Measuring rejection is non-trivial because suggestions may be modified or deleted after they are initially accepted. We therefore distinguish between different behaviors that result in partial or complete rejection of a suggestion.

\begin{itemize}
    \item[\textbf{AM}] \textit{Accept\_then\_Modify}: Editing a suggestion immediately after \textit{tab\_accepting} it. This includes removing characters (but not all the characters). This behavior has been observed but not named~\cite{jayagopal2022learnability}, or previously referred to as \textit{accept and adapt}~\cite{prather2023weird}, \textit{accept and edit}~\cite{barke2023grounded}, and \textit{modify}~\cite{kazemitabaar2024selfpaced}. There are also related metrics, such as \textit{accept\_then\_repair}~\cite{vaithilingam2022expectation}, which have the additional connotation of correctness.
    \item[\textbf{AD}] \textit{Accept\_then\_Delete}: Deleting a suggestion immediately after \textit{tab\_accepting} it. This behavior was observed qualitatively as \textit{backtracking}~\cite{prather2023weird}. A similar behavior, \textit{delete and search}, involves attempted debugging before deletion~\cite{vaithilingam2022expectation}.  
    \item[\textbf{I}] \textit{Ignore}: Dismissing a suggestion without directly interacting with it. This could include running code, typing something else, or clicking elsewhere.

\end{itemize}

\subsubsection{Execution Events} We also tracked when students ran their code. Running code is consistently associated with higher task performance~\cite{carterNormalized2015}, and so this metric was important to include: 
 
\begin{itemize}
    \item[\textbf{RC}] \textit{Run\_Code}: This captures when students compile and run their code, and shows the tests that pass and fail.
    \item[\textbf{NR}] \textit{Number\_of\_Runs} Total number of times student ran their code throughout entire session. $\text{NR} = \text{count}(RC)$ 
\end{itemize}

\subsubsection{Derived Metrics} From the event-level logs, we derived per-suggestion metrics that characterize participants' engagement and decision-making processes. These include when users \textbf{Accept} code without modification, \textbf{Reject} code by interacting with the system in any way other than accepting the suggestion, \textbf{Fail an Attention Check} by accepting a deliberately incorrect line of code, or the \textbf{Dwell Time} which is the time \textit{measured in milliseconds} between when a suggestion is shown to the \textit{start} of any user action.

\begin{itemize}
    \item[\textbf{A}] \textit{Accept} $= TA \lor SA$ 
    \item[\textbf{R}] \textit{Reject} $= AM \lor AD \lor I$ 
    \item[\textbf{FC}] \textit{Failed Attention Check} $=(TA \lor SA) \land (GAC \land SS) $
    \item[\textbf{DT}] \textit{Dwell\_Time}  $= t_{\text{action}} - t_{SS}, \quad
        t_{\text{action}} \in \{TA, SA, AM, AD, I\}$

\end{itemize}

\subsubsection{Aggregated Session Metrics}

Based on the metrics described in the previous section, we also produce metrics that described more complex behaviors and aggregated behavior across the session. 

\begin{itemize}

    \item[\textbf{AR}] \textit{Acceptance\_Rate} 
    $ =  (count(TA) + count(SA)) / count(SS)$
    
    \item[\textbf{TAR}] \textit{Tab\_Acceptance\_Rate} 
    $= count(TA)/count(SS)$
    \item[\textbf{SAR}] \textit{Slow\_Acceptance\_Rate} 
    $ = count(SA)/count(SS)$
    
    \item[\textbf{AMR}] \textit{Accept\_then\_Modify\_Rate} 
    $ = count(AM)/count(SS)$

    \item[\textbf{ADR}] \textit{Accept\_then\_Delete\_Rate} 
    $ = count(AD) / count(SS) $

    \item[\textbf{FCR}] \textit{Failed\_AC\_Rate} 
    $= count(FC)/count(GAC \land SS)$

    \item[\textbf{RR}] \textit{Reject\_Rate} 
    $= count(R) / count(SS)$
    
    \item[\textbf{ADT}] \textit{Average\_Dwell\_Time} 
    $ = \sum DT / count(SS)$
\end{itemize}

\subsubsection{Outcome Metrics} We also measured task-level outcomes to assess overall solution correctness.

\begin{itemize}
    \item[\textbf{TP}] \textit{Task\_Performance}: Total number of test cases passed out of all possible test cases for programming problem.
    
\end{itemize}

\subsection{Correlation Analysis} 

After computing the metrics presented in the previous section, we summarized these values using box plots to show their distribution. To understand the relationships between the metrics, we computed a Spearman correlation matrix using session-level aggregates. To ensure a consistent level of abstraction, all event-derived measures were first aggregated at the session level prior to analysis to ensure a consistent unit of analysis across variables.
Our analyses focused on examining the relationships between session-level behavioral measures, which include dwell time, acceptance rate, slow acceptance rate, accept-then-modify rate, and accept-then-delete rate, and outcome measures of task performance and incorrect suggestion acceptance (failed attention checks).

\section{Results}
\begin{figure}
    \centering
    \includegraphics[width=\linewidth]{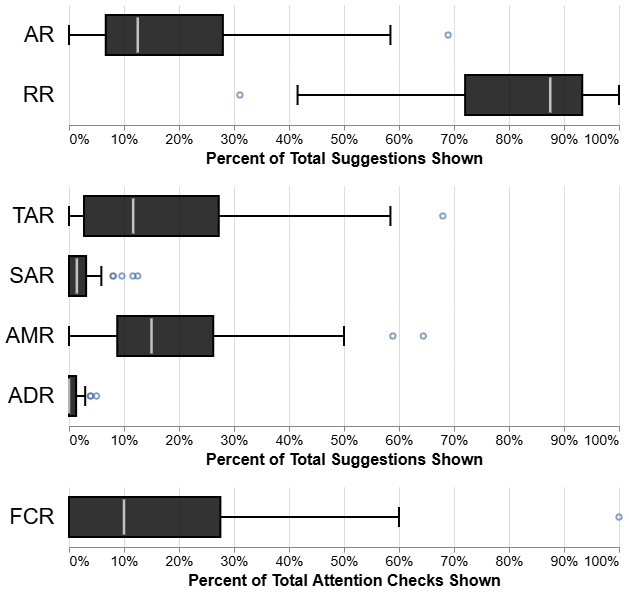}
    \caption{Distribution of session-level interaction rates.}
    \label{fig:distributions}
\end{figure}

\begin{figure}
    \centering
    \includegraphics[width=\linewidth]{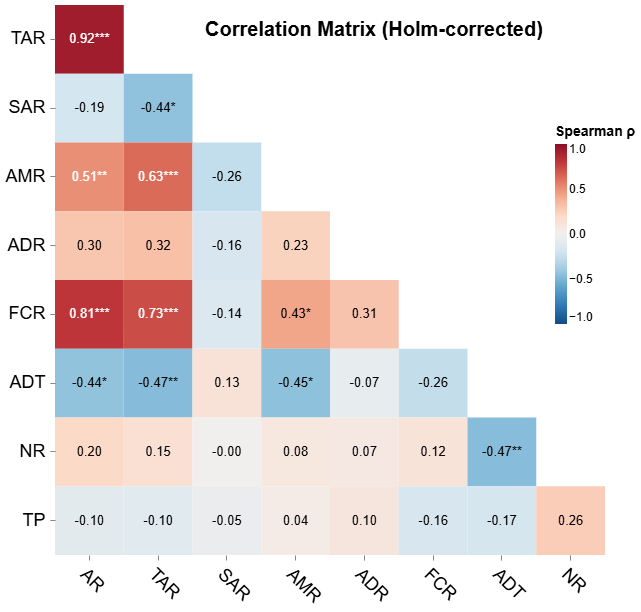}
    \caption{Correlations between session-level metrics using Spearman's Rho. (* p<.05  ** p<.01  *** p<.001) }
    \label{fig:correlations}
\end{figure}

Figure~\ref{fig:distributions} presents box plots that summarize the distribution of behavioral metrics. We observed that participants \textbf{accepted (A)} an average of 18.4 suggestions ($SD=30.9$). Participants \textbf{tab\_accepted (TA)} an average of 17.1 suggestions ($SD=31.0$), and 45 participants \textbf{TA} at least once. Participants \textbf{slow\_accepted (SA)} an average of 1.3 times ($SD=1.5$), and 34 participants \textbf{SA} at least once. On average, participants \textbf{accept\_then\_modified (AM)} 17.7 times ($SD=29.8$), and 49 participants \textbf{AM} at least once. Participants \textbf{accept\_then\_deleted (AD)} an average of 0.6 times ($SD=1.0$), and 19 participants \textbf{AD} at least once. Participants \textbf{ignored (I)} an average of 36.3 suggestions ($SD=13.3$), and all participants \textbf{I} suggestions. \textbf{Average\_dwell\_time (ADT)} was 12.2 seconds ($SD=8.1$), the minimum was 1.7 seconds, and the maximum was 39.1 seconds. 
We analyzed session-level metrics, including \textbf{number\_of\_runs (NR)} and \textbf{task\_performance (TP)}.  The average \textbf{NR} was 15.5 ($SD=12.6$), the minimum was 0, and the high was 52. 
The average \textbf{TP} was 10.4 tests ($SD=12.9$), the minimum was 0, and the maximum was 26. Out of the 55 participants, 22 students successfully completed the task, passing all 26 test cases.

\subsection{RQ1: Interactions and Task Performance}

As shown in Figure~\ref{fig:correlations}, interaction metrics were weakly or inconsistently correlated with task performance. Code execution, dwell time, and failed attention checks were most strongly correlated though they remained moderate to weak.

\subsubsection{Code Execution (Number of Runs)}

Among the examined metrics, \textbf{NR} showed the strongest positive correlation with \textbf{TP} ($\rho=0.26$), although the relationship was modest. This finding aligns with previous studies, which suggest that running code is often associated with higher task performance~\cite{carterNormalized2015}. However, \textbf{NR} was also negatively correlated ($\rho=-0.47$) with \textbf{ADT}. The longer students look at suggestions before acting, the less they ran their code. One possible explanation is that some students quickly accepted suggestions with minimal engagement only to use \textbf{NR} to verify the logic.

\subsubsection{Average Dwell Time}

\textbf{ADT} negatively correlated ($\rho=-0.26$) with \textbf{FCR}, which could mean the less time students spend on a suggestion, the more likely they are to have high \textbf{FCR}. This suggests that students who spent more time looking at a suggestion passed more attention checks. However, among all the metrics we logged, \textbf{ADT} had the strongest negative correlation ($\rho=-0.17$) with \textbf{TP}, suggesting that the less time students spend on a suggestion, the higher their \textbf{TP}. This indicates that \textbf{DT} does not seem to directly relate to critical engagement.

\subsubsection{Takeaway:} These findings suggest that there is no perfect measure of responsible AI use. Even behaviors often associated with more expert use, such as \textbf{NR}, were positively correlated with \textbf{FCR}, indicating some lapses in critical engagement. 

\subsection{RQ2: Attention Checks and Performance}

While no clear metric was strongly correlated with task performance, several metrics were correlated with unreflective engagement with suggestions (i.e.: failed attention checks).

\subsubsection{Tab Accept Rate}

\textbf{TAR} weakly negatively correlated ($\rho=-0.10$) with \textbf{TP}, suggesting that students with high \textbf{TAR} were slightly less likely to perform well. Furthermore, students who \textbf{TAR} were far more likely to fail attention checks ($\rho=0.73, p<0.001$). \textbf{TAR} was also strongly negatively correlated with \textbf{ADT}, ($\rho=-0.47$) 
indicating that students with high \textbf{TAR} tended to have lower \textbf{DT}.

\subsubsection{Attention Checks and Task Performance}

We investigated two measures of misuse: \textbf{TP} represents an implicit measure of potential misuse, whereas \textbf{FCR} offers a more explicit measure. \textbf{FCR} and \textbf{TP} had a weak negative correlation ($\rho=-0.17$). This suggests that students who passed attention checks did not necessarily perform better on the task. Therefore, \textbf{FCR} does not represent critical engagement, but it does suggest potential misuse.

\subsubsection{Takeaway:} While behaviors and attention checks were not strongly correlated with performance, we observed that some behaviors, such as tab accept were strongly correlated with failed attention checks, suggesting a potential lapse in reflective engagement with suggestions.

\subsection{Novel Behavioral Metrics}

Our log analysis also identified new behavioral metrics including: 

\begin{description}
    \item[\textbf{STA}] \textit{Slow Tab Accept:} Starting a \textbf{SA} by typing out the suggestion manually, clicking `tab' before completing the \textbf{SA}.  
    \item[\textbf{SM}] \textit{Slow Accept then Modify:} Completing the \textbf{SA}, but then immediately modifying the suggestion.  
    \item[\textbf{SD}] \textit{Slow Accept then Delete:} Completing a \textbf{SA}, but then deleting all or some of the suggestion. 
\end{description}

These metrics were somewhat common in our dataset. 
Of the 71 \textbf{SA} events, we observed 10 \textbf{STA} events and 53 \textbf{SM} events. \textbf{STA} and \textbf{SM} accounted for 75\% of our \textbf{SA}, making it a majority. \textbf{SD} was not observed in our log data, but logically follows from the possibility of \textbf{SM}. 

These metrics highlight that a lot can happen in the process of accepting a suggestion. With the inclusion of these metrics, our taxonomy becomes even more robust in capturing how students interact with code suggestions. 

These behaviors indicate that suggestion acceptance is not a terminal action, but often part of a multi-step editing process involving revision and correction. Suggestions have a life that extends beyond when they are accepted as students may return to them at various points and make modifications or delete them.

\color{black}\section{Discussion}

Across recent studies of AI code completion tools, students' behaviors and strategies have been observed qualitatively~\cite{barke2023grounded, prather2024widening, prather2023weird}. Our work extends this literature by operationalizing these behaviors into a taxonomy of measurable interaction metrics and demonstrates how these metrics can be captured quantitatively at scale using Clover. Consistent with these previous qualitative findings, we observed substantial diversity in how students engage with AI code suggestions. 
Our findings also suggest that while metrics are weakly or inconsistently correlated with performance, some behaviors tend to be more consistently associated with unreflective engagement with suggestions (i.e.: failed attention checks).

\subsection{Interaction Metrics for AI Code Suggestions}

\subsubsection{Interactions patterns varied across participants}

We observed variation across all behavioral metrics including acceptance behaviors, dwell time, code execution frequency, and post-acceptance revision behaviors. This heterogeneity aligns with insights from learning sciences that describe novice learning as non-linear and diverse in strategies, rather than a stable progression through stages of expertise. Students may shift fluidly between exploratory and exploitative strategies to solve the problem at hand. This variability has already been well documented in recent studies of AI code completion tools~\cite{shihab2025effects, prather2024widening}.

These metrics also have the potential to operationalize multiple aspects. For example, average dwell time (\textbf{ADT}) might relate to both cognitive processing and expertise. From dual-process theory~\cite{kahneman2011thinking}, shorter dwell times may indicate rapid, intuitive judgments, while longer dwell times may indicate more deliberate reflective reasoning. Alternatively, considering expertise, short dwell times may reflect negative expertise~\cite{minsky1997negative}, where experienced students rapidly recognize a low-quality suggestion and reject it. Novices may spend more time attempting to interpret the suggestion, or may engage superficially without meaningfully evaluating it.

\subsubsection{Interaction Metrics and Task Performance}

Across the behavioral metrics studied, we did not observe a strong or consistent relationship with task performance. While several behaviors which are traditionally associated with productive debugging and learning, such as running code, had modest associations with performance, these relationships were generally weak or inconsistent. This suggests that behaviors typically associated with good performance may coexist with lapses in critical engagement.

\subsubsection{Interaction Metrics and Failed Attention Checks}

In contrast, several behaviors showed more consistent patterns associated with potential lack of reflective engagement with suggestions. For example, high rates of \textbf{tab acceptance} were strongly associated with failed attention checks. This relationship was substantially stronger than any observed association between other interaction metrics and task performance. Slow accept was the only behavior negatively correlated with failed attention checks. \textbf{Dwell time} had a weaker but consistent pattern. Longer dwell times were associated with better attention check performance and slightly improved task performance. This suggests that looking more carefully at suggestions tends to improve performance. These findings suggest that interaction patterns, such as frequent tab acceptance and to a lesser extent shorter dwell times, may serve as indicators of reduced engagement with AI-generated code suggestions.

\subsection {A Taxonomy of Interaction Metrics}

As shown in Section~\ref{sec:taxonomy}, there has been a need to systematize the behavioral metrics associated with AI code completion tools. These metrics often used different terms or were measured differently across studies. Our work contributes to this goal by formalizing a taxonomy of metrics capturing how students engage with AI code completion tools. This paper offers an example of how this taxonomy might be used to better understand student-AI interactions, but future work could more extensively interrogate sequential behaviors and patterns of interaction over time.

\subsection{Limitations} 

This study faces a number of limitations, which are appropriate given our goal to develop and prototype a taxonomy of behavioral metrics for how students use AI code completion tools.

\subsubsection{Single Problem and Task Performance} While the rainfall problem used in this study has been commonly studied in computing education research, future work should investigate multiple diverse problem types. Task performance on the problem contained limited variability, with some participants producing non-compiling solutions and others receiving fully correct outputs which resulted in a lack of intermediate performance levels.

\subsubsection{Single Model and Latency} Our system relied on a single AI model, but there are significant differences in performance and latency based on the model selection.

\subsubsection{Post-Accept Behaviors} We measure dwell time by calculating the time between the suggestion being shown and the users's first interaction. However, students may continue engaging with suggestions long after they have been accepted. These longer-term behaviors were not analyzed in this study. 

\subsubsection{Participant Sample} 

Participants varied in their experience with programming and with GitHub Copilot, but they only represent the students from a single class within a single university. 

\subsubsection{Potential for Cheating}
This study was conducted in a classroom setting, and students received participation credit that did not depend on their performance. Instructions also emphasized the importance of genuine effort, but we cannot rule out the use of external tools such as Google or ChatGPT.

\subsubsection{Suggestion Types and Attention Checks} 

Clover's attention checks injected deterministic incorrect suggestions to probe critical engagement. While this approach allows controlled measurement of misuse, it does not fully replicate spontaneous AI hallucinations in real-world coding. Moreover, the study did not account for the type or complexity of suggestions students received, which may influence acceptance, modification, or rejection behaviors.

\subsection{Future Work}

Our work opens a number of potential research threads. 

\subsubsection{Suggestion Lifecycles} 

Our work focuses on atomic interaction events, but students may continue to interact with suggestions long after they are accepted. Future work could investigate these longer-term interactions with suggestions. 

\subsubsection{Partial Accepts}

Our current implementation of slow accept only includes \textit{exact} matches to code suggestions. Future work could explore partial matches by measuring the similarity between the suggested and accepted code using Levenshtein distance and cosine similarity. This will provide a more nuanced characterization of slow accepts and whether they reflect verification, partial reuse, or adoption of a new approach. 

\subsubsection{Attention Checks} 

In this study, attention checks were used as a behavioral probe to assess students' engagement with code suggestions. Future work could explore shifting attention checks from a passive metric toward a feedback mechanism. Students could receive real-time or summative feedback about how they engage with the attention checks.

\subsubsection{Incorporating Student Demographics}

Prior work suggests that programming background, AI literacy, and self-efficacy may influence how students interact with AI assistants~\cite{margulieux2024self,prather2024widening}. Future studies should investigate how learner characteristics and demographics affect interaction patterns and behaviors. Incorporating demographic information with interaction logs could explain variations and patterns in acceptance strategies, rejection behavior, and engagement for a diverse student population. 

\subsubsection{Longitudinal Studies} Our current study analyzes interactions within a single task, but over time, students' behaviors may change. Studying interactions across multiple assignments or throughout the duration of a course could reveal fluctuations in students engagement with AI-generated suggestions and determine the effect of problem type on students' behaviors.

\section{Conclusion}

In this work, we introduced Clover, an AI code completion tool that captures behavioral metrics and uses attention checks to probe critical engagement. Our study with CS1 students shows that engagement with code suggestions is highly variable. Metrics may reflect a complex mix of reflective reasoning, over-reliance, or negative expertise. As a result, no single metric reliably captures critical engagement. By formalizing a taxonomy of behavioral metrics, we provide a framework for interpreting patterns of student-AI interaction which can create a path for future research.

\balance

\bibliographystyle{ACM-Reference-Format}
\bibliography{base}

\end{document}